\documentclass[10pt]{iopart}
\usepackage{graphicx}
\usepackage{epsfig}
\newcommand{\tw}{t_\mathrm{w}}

\begin{document}

\title[Aging and ultra-slow equilibration in concentrated colloidal hard
spheres]{Aging and ultra-slow equilibration in concentrated colloidal hard
spheres}

\author{D. El Masri, M. Pierno, L. Berthier,
and L. Cipelletti\footnote[3]{To whom correspondence should be addressed:
lucacip@lcvn.univ-montp2.fr}}

\address{Laboratoire des Collo\"{\i}des, Verres et Nanomat\'eriaux,
UMR 5587 CNRS and Universit\'{e} Montpellier II,
34095 Montpellier Cedex 5, France}

\begin{abstract}
We study the dynamic behaviour of concentrated colloidal hard
spheres using Time Resolved Correlation, a light scattering
technique that can detect the slow evolution of the dynamics in
out-of-equilibrium systems. Surprisingly, equilibrium is reached a
very long time after sample initialization, the non-stationary
regime lasting up to three orders of magnitude more than the
relaxation time of the system. Before reaching equilibrium, the
system displays unusual aging behaviour. The intermediate
scattering function decays faster than exponentially and its
relaxation time evolves non-monotonically with sample age.
\end{abstract}

\section{Introduction}
\label{Intro}

In the last two decades, colloidal hard spheres have been widely
studied as a model system that exhibits a transition to an
arrested glassy state as particle volume fraction, $\varphi$,
increases \cite{PuseyNature1986}. Above $\varphi \approx 0.49$,
the dynamics dramatically slows down and exhibits a
two-step relaxation, where the initial relaxation corresponds to
the motion of the particles within the cage formed by their
neighbours, while the final, $\alpha$-relaxation is due to
cage-escape processes~\cite{vm,vanMegenPRE1998}.
Probably, the most popular experimental
technique for investigating these dynamics has been dynamic light
scattering, which has allowed the microscopic dynamics of samples
at thermal equilibrium to be probed for volume fractions up to
$\varphi \approx 0.55$~\cite{vanMegenPRE1998}.
Light scattering experiments
measure the time autocorrelation function of the fluctuations of
the scattered intensity, $g_2(\tau)-1$, which is proportional to
the squared intermediate scattering function \cite{Pecora}.
Confocal microscopy was used in more recent works that focused on
dynamic heterogeneity, identifying localized clusters of particles
whose dynamics differ from the 
average~\cite{WeeksScience2000,KegelScience2000}.

By contrast, experiments probing the non-equilibrium or aging
behaviour of concentrated samples are scarce
\cite{vanMegenPRE1998,PhamPRE2004,CourtlandJPCM2003,SimeonovaPRL2004}.
This is because conventional light scattering techniques require a
considerable time averaging to get sufficient statistics. Clearly,
extensive time averaging prevents aging dynamics to be precisely
characterised in systems where the dynamics are not stationary.
Microscopy experiments can in principle provide time-resolved
information, but they have often very limited statistics, since
they are forced to deal with rather large particles, for which the
dynamics is too slow to be followed in the aging regime over
several decades.

In this work, we take advantage of recently-introduced dynamic
light scattering techniques,
multispeckle~\cite{BartschJChemPhys1997,LucaRSI1999} and Time
Resolved Correlation (TRC)~\cite{LucaJPCM2003}, to follow the slow
evolution of the dynamics of concentrated suspensions of colloidal
hard spheres, from an initial aging regime up to times where
thermal equilibrium is reached. We find that hard spheres display
an aging behaviour that is quite different from the one observed
in molecular supercooled liquids. Strikingly, thermal equilibrium
is reached only for times that considerably exceeds the relation
time of the system and an unusual `compressed' exponential
relaxation of $g_2-1$ is observed in the aging regime.
Additionally, we find that in the stationary regime the dynamics
appears to be extremely sensitive to a small shear, an issue
possibly relevant when devising experimental methods dealing with
very concentrated samples.

\section{Experimental methods}
\label{sec:exp}

\subsection{Experimental setup}
\label{sec:setup}

\begin{figure}
\begin{center}
\psfig{file=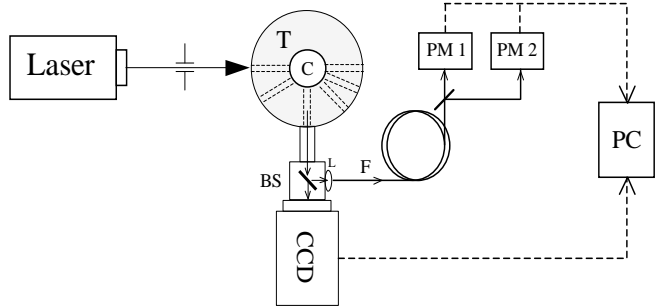,width=9cm,height=4.5cm} \caption{\label{Fig1}
Sketch of the experimental setup described in
section~\ref{sec:setup}.}
\end{center}
\end{figure}

The experimental setup is shown in \fref{Fig1}. The light source
is a frequency--doubled Nd:YAG laser whose vertically polarized
beam impinges on a cylindric scattering cell, C. The cell
temperature is set to $T = 20 \pm 0.05 ^{\circ}{\rm C}$ via water
circulating in a copper cell holder, T. Light scattered within a
volume of about 1 ${\rm mm}^3$ at a scattering angle
$\theta=90^{\circ}$ (which corresponds to wavevectors close to the 
first peak of the structure factor of our hard sphere system)
is split by the beam splitter BS so as to
illuminate a charge-coupled device (CCD) detector and to be
collected by a single-mode, polarization-maintaining fiber optics.
Two photomultipliers connected to a hardware correlator hosted in
a personal computer (PC) allow the fiber optics signal to be
analyzed to obtain $g_2(\tau)-1$ down to a lag time $\tau = 50$
nsec, for samples whose relaxation time does not exceed $100$ sec
and whose dynamics are stationary (here, $g_2-1$ is obtained by
extensive time-averaging).

The CCD images of the speckle pattern generated by the sample are
transferred to the PC via a frame grabber and stored on the hard
drive for later processing via a software correlator. The quantity
measured in our TRC experiments is the degree of correlation,
$c_I(t,\tau)$, of the scattered intensity of two images separated
by a lag $\tau$~\cite{LucaJPCM2003}:
\begin{equation}
\label{cIdef}
c_I(t, \tau) = \frac {\langle I_p \, (t) \; I_p \, (t + \tau)
\rangle_{p}} {\langle I_p \, (t) \rangle_{p} \; \langle
I_p \, (t + \tau) \rangle_{p}} -1,
\end{equation}
where $I_p(t)$ is the instantaneous value of the scattered
intensity measured by the $p$-th pixel at time $t$, and the
average $\langle ... \rangle_{p}$ is taken over all CCD pixels.
Note that no time average is performed; thus, the degree of
correlation depends on both the time lag $\tau$ between the two
images analyzed and the time $t$ at which the first image of the
pair is taken. In systems with stationary dynamics a further
average over $t$ can be performed. The normalized
auto--correlation function is then obtained
\begin{equation}\label{g2norm}
g_2(\tau) -1 = \langle  c_I(t, \tau)  \rangle_{t},
\end{equation}
similarly to conventional dynamic light scattering techniques. The
minimum time lag for which $c_I$ can be calculated, $\tau \simeq
20$~ms, is limited by the speed of the CCD camera and the exposure
time. By coupling DLS data obtained from the hardware correlator
to TRC, we are able to measure simultaneously
and on the same sample the intensity correlation function from
$\tau=50$~ns up to very large lag times, $\tau \simeq 10^4$~s.

\subsection{Sample preparation}

The particles used in this work are poly-methylmethacrylate
(PMMA) spheres of radius $R=140$ nm, with a degree of
polydispersity of about 10\% that prevents crystallization in the
experimental timescale. The particles are sterically
stabilized by a thin layer of poly-12-hydroxystearic acid of
approximately 10 nm thickness. They are suspended in a mixture of
organic solvents (cis-decalin and tetralin) that closely matches
their refractive index and reduces van der Waals 
interactions.

Samples are prepared by centrifugation of a stock suspension for
about 24 h at 1000~$g$ to obtain a dense amorphous sediment,
assumed to be near random close packing, $\varphi_0 \approx 0.67$,
for a 10\% polydispersity~\cite{sillescu}.
The clear supernatant is removed and a
controlled mass of solvent is added to obtain the desired volume
fraction, $\varphi$. Samples are then dispersed by
vortexing and tumbling for about 6 h. Before starting
measurements the sample is thermalized at $20^{\circ}$C for 15
minutes. In aging experiments, the waiting time, $\tw$, is taken
from the end of the dispersion, which corresponds to
$\tw=0$. Since our sample is very polydisperse we cannot 
use the freezing volume fraction to precisely control volume 
fractions~\cite{PuseyNature1986,vm,vanMegenPRE1998}, 
leading to rather large uncertainties on the calculated $\varphi$. 
Indeed, a comparison of our dynamic data with the 
literature~\cite{PuseyNature1986,vm,vanMegenPRE1998} 
suggests that we systematically underestimate volume fractions
by 2 to 4\%. This should be kept in mind when comparing the 
present results to the literature.

\section{Results}
\label{sec:results}

\begin{figure}
\psfig{file=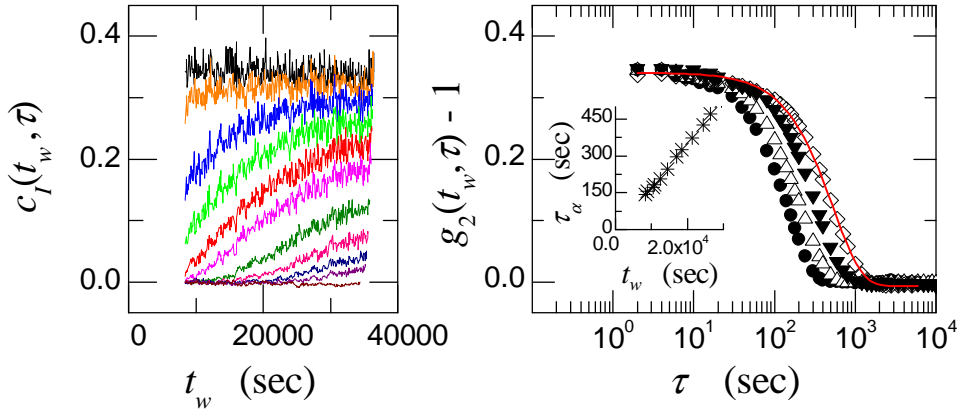,width=13cm} \caption{\label{Fig:agingsample}
Left: age dependence of the degree of correlation,
$c_I(\tw,\tau)$, measured for a sample with $\varphi = 0.545$. From
bottom to top, $\tau=2000$, 1200, 1000, 800, 600, 400, 300, 200, 120,
40, 2 sec. The steady growth of $c_I(\tw,\tau)$ with $\tw$
indicates aging.
Right panel: time decay 
of $g_2(\tw,\tau)-1$, obtained by averaging $c_I(\tw,\tau)$
over a time window of a few hundreds sec, for 
$\tw = 8400$, 12400, 18400, and 29400 sec (from left to right).
The shape of the relaxation is well described by a `compressed'
exponential form, as shown by the full line which has
a streching exponent $p = 1.22$. 
Inset: Linear dependence of the $\alpha$-relaxation time
on $\tw$ typical of a `simple aging' behaviour.}
\end{figure}

The left panel of \fref{Fig:agingsample} shows the waiting time
evolution of the degree of correlation $c_I(\tw,\tau)$ measured
for a sample at $\varphi = 0.545$, for various time lags $\tau$
ranging from 2 to 2000 sec and corresponding to the 
$\alpha$-relaxation of the system. At fixed lag, $c_I$ continuously
increases with $\tw$, demonstrating that changes of the sample
configuration during a fixed time interval become smaller when
$\tw$ increases. In other words, the dynamics slows down when the
system gets older, a clear manifestation of
aging~\cite{Bouchaud2000}. The right panel shows a set of
representative intensity correlation functions $g_2(\tw,\tau)-1$
obtained by averaging $c_I$ over a short time window (a few hundreds of sec),
starting at various $\tw$.
To extract the characteristic time, 
$\tau_{\alpha}(\tw)$, of the $\alpha$-relaxation,
we fit the curves by a stretched exponential: $g_2(\tw,\tau)-1 = a
\exp\left [-(\tau/\tau_{\alpha})^p\right]$. The inset shows
$\tau_{\alpha}$ vs. $\tw$, indicating that the relaxation time
increases linearly with sample age, a behaviour observed in many
glassy systems \cite{Bouchaud2000} and refereed to as `simple
aging'. Note however that $\tau_{\alpha}/\tw \ll 1$, thus ruling
out the simple explanation stating that the relaxation time is set
by the age, $\tau_{\alpha} \approx \tw$. Surprisingly, we find
that for all curves $p \approx 1.3$. This behaviour is in contrast
with that of concentrated colloidal hard spheres at equilibrium,
where $p < 1$ \cite{vm}. Intriguingly, it is
similar to the `compressed' exponential relaxation found in a
variety of soft glassy materials out of 
equilibrium~\cite{CipellettiJPCM2005}.

\begin{figure}
\begin{center}
\psfig{file=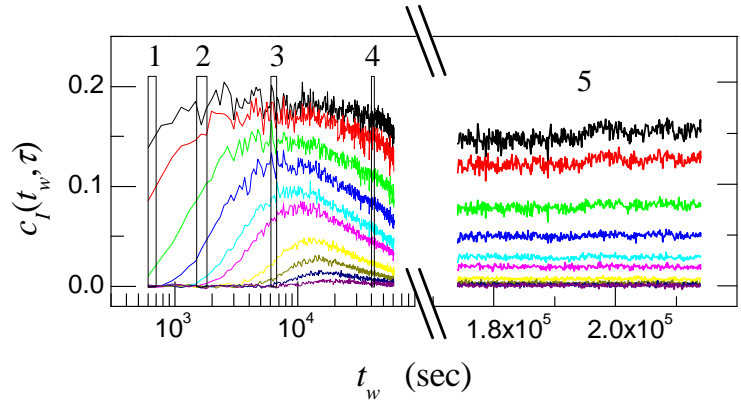,width=10cm}
\end{center}
\caption{\label{Fig:cImorediluted} Degree of correlation as a
function of $\tw$ for a sample with $\varphi = 0.505$. From
bottom to top, $\tau = 200$, 140, 100, 70, 40, 30, 18, 10, 4, 2
sec. Before the $x$ axis break, the scale is logarithmic, in order
to better appreciate the initial evolution of the dynamics. 
The rectangles indicate the time
windows over which $c_I$ is averaged to obtain $g_2 - 1$ shown in
\fref{Fig:g2morediluted}.}
\end{figure}

The sample shown in \fref{Fig:agingsample} does not reach
equilibrium on the experimental time scale. In order to
investigate how a stationary state is eventually reached, we turn
to a more diluted sample, $\varphi = 0.505$.
\Fref{Fig:cImorediluted} shows the $\tw$ dependence of $c_I$. A
striking behaviour is observed. Initially, $c_I$ grows, as
also observed in \fref{Fig:agingsample}.
However, after reaching a maximum, the
degree of correlation then decreases with $\tw$, before eventually
reaching a stationary regime. This implies that after the initial
simple aging regime, the dynamics accelerates and finally
becomes stationary. Moreover, for different $\tau$ the maximum of
$c_I$ occurs at different ages, implying that also the shape of $g_2-1$
changes with time.

\begin{figure}
\begin{center}
\psfig{file=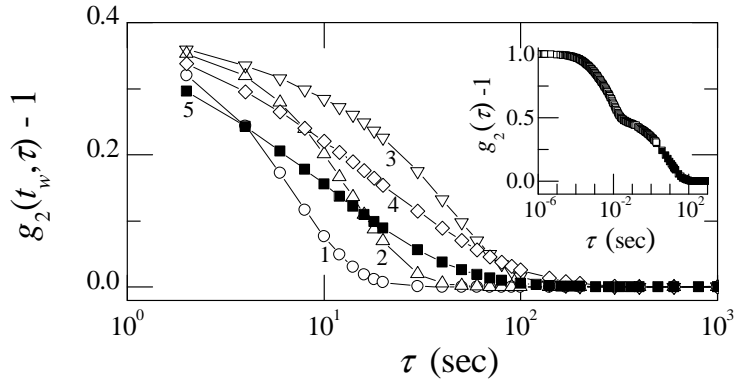,width=10cm}
\end{center}
\caption{\label{Fig:g2morediluted} Intensity
correlation functions calculated from the $c_I$ of
\fref{Fig:cImorediluted}. Each curve is labelled by the
time window shown in \fref{Fig:cImorediluted}. 
The initial aging regime ($1\to2\to3$) is followed
by an acceleration of the dynamics ($3\to4\to5$)
and finally by a stationary
regime (5).
Inset, CCD data
(solid squares) and hardware correlator data (open squares)
collected simultaneously in the stationary regime.}
\end{figure}

The evolution of $g_2-1$ is shown in \fref{Fig:g2morediluted}.
Initially the relaxation time increases and the correlation
functions are compressed ($p \approx 1.3$). Then, $\tau_{\alpha}$
goes back to smaller values. Concomitantly, the shape changes and
becomes more stretched. For $\tw \geq 1.8 \times 10^5$ sec the
dynamics becomes stationary. Thus, equilibrium is reached after
more than 1000 relaxation times, a behaviour at odd with that of
molecular glassy systems \cite{Bouchaud2000}. This
suggests that motion on length scales much larger than the particle
size might be involved in this process. At equilibrium, we
measure the fast dynamics with the hardware correlator by
averaging $g_2-1$ over 12 h. The inset of \fref{Fig:g2morediluted}
shows the full $g_2-1$ obtained by pasting together hardware
correlator and CCD data. A typical two-step decay with a final
stretched relaxation ($p \approx 0.6$) is observed.

\begin{figure}
\begin{center}
\psfig{file=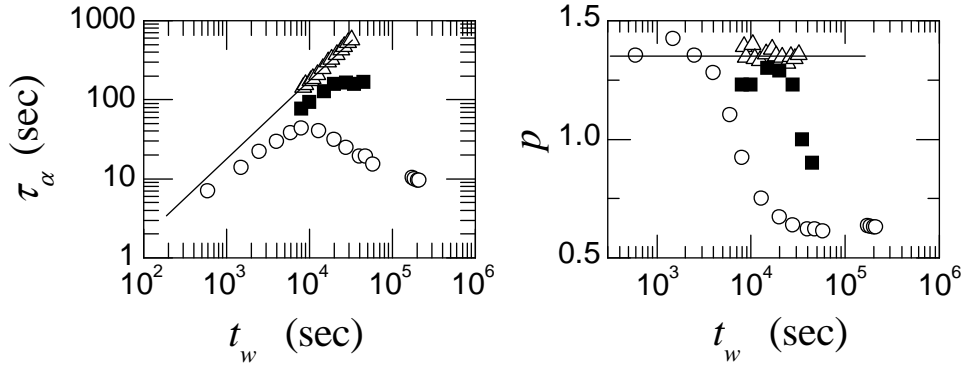,width=13cm}
\end{center}
\caption{\label{Fig:agedependence}Age evolution of the relaxation
time $\tau_{\alpha}$ (left panel) and the stretching exponent $p$
(right panel) for samples at $\varphi = 0.545$ (open triangles),
$\varphi = 0.54$ (solid squares), and $\varphi = 0.505$ (open
circles). The lines $\tau_\alpha \propto \tw$ and $p=1.3$ 
are guides to the eye.}
\end{figure}

Values of the $\tw$ dependence of $\tau_{\alpha}$ and $p$ are
shown in \fref{Fig:agedependence} for the two samples discussed
above and for an additional intermediate volume fraction. For the
lower $\varphi$ (open circles), three regimes are observed: aging
($\tau_{\alpha} \propto \tw$ and $p \approx 1.3$), acceleration of
the dynamics ($\tau_{\alpha}$ and $p$ decrease with $\tw$),
stationary dynamics ($\tau_{\alpha}$ and $p$ constant). As the
volume fraction increases, the end of the aging regime shifts
towards larger $\tw$, eventually becoming too large to be
detected.

\begin{figure}
\begin{center}
\psfig{file=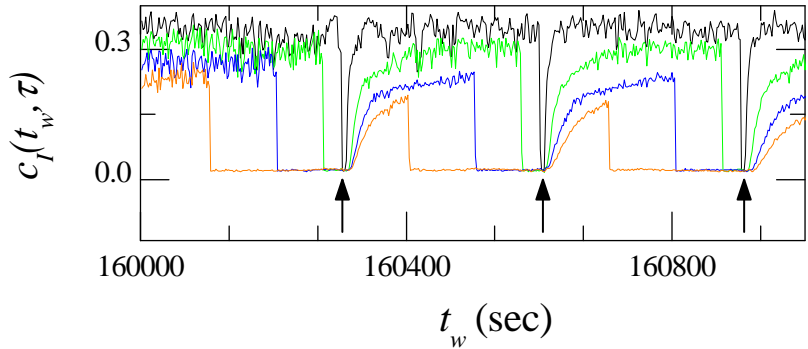,width=11cm}
\end{center}
\caption{\label{Fig:turning} Time evolution of the degree of
correlation for a sample with $\varphi = 0.496$, after reaching
equilibrium (from bottom to top, $\tau=200$, 100, 30, 2 sec). The
sample is slightly rotated at three different times indicated by the
arrows. Recovery after each rotation is very slow.}
\end{figure}

Our findings indicate that it takes a surprisingly long time for
concentrated hard spheres to recover from a mechanical
perturbation, such as the initial homogenization. This
equilibration time may become prohibitively long when
$\tau_{\alpha}$ exceeds 1000 s. Moreover, ultra-slow equilibration
is potentially relevant to several measuring schemes used to deal
with glassy dynamics, where the sample cell is rotated to
accumulate statistics, either in between short DLS runs when
measuring short time dynamics (`brute force' method
\cite{PuseyPhysicaA1989}), or continuously to access the slow
dynamics (`interleaved' \cite{MullerProgColloidPolymSci1996} or `echo'
\cite{PhamRSI2004} methods). Indeed we show the effect of a tiny
rotation of the cell in \fref{Fig:turning}. Initially, the degree
of correlation is almost constant, since the sample is at
equilibrium. When the sample is turned manually by the smallest
possible amount (arrows in the figure), $c_I$ drops to 0, because
the speckle pattern changes completely. Strikingly, it takes a
very long time for $c_I$ to grow back to its initial value.
Moreover, the larger $\tau$ the longer the recovery time. We note
that at the smallest delay shown in \fref{Fig:turning}, $c_I$
stabilizes in a few seconds, suggesting that the short time
dynamics should not be significantly affected by the sample
rotation. By contrast, on the longer time scales probed by the
interleaved or echo methods, sample rotation may spuriously
accelerate the dynamics. It would be interesting to also test these
effects in setups where sample rotation is
done by a stepper motor.

We thank A. Schofield, W. Poon, P. Pusey, Q. Pham, and G.
Petekidis for providing the colloidal particles and for helpful
discussions. This work is supported by the European MCRTN
``Arrested matter'' (MRTN-CT-2003-504712), the NoE ``Softcomp'',
CNES, and the French Research Ministry (ACI JC2076). L.C. is a
junior member of the Institut Universitaire de France who also
supports his research.

\end{document}